\documentclass[aip,reprint, eprintnumbers, preprintnumbers]{revtex4-2}
\usepackage[]{graphicx}
\usepackage{epstopdf}
\usepackage{bm}
\usepackage{hyperref}
\usepackage{framed}
\usepackage{color}
\newcommand{\vo}{V$_2$O$_3\,$}

\begin{document}
\title{Absolute calibration of the latent heat of transition using differential thermal analysis}
\author{Tapas Bar}\email{tapasbar1993@gmail.com}
\affiliation{Indian Institute of Science Education \& Research Kolkata, Mohanpur Campus, Nadia 741246, West Bengal, India}
\author{Bhavtosh Bansal}\email{bhavtosh@iiserkol.ac.in (corresponding author)}
\affiliation{Indian Institute of Science Education \& Research Kolkata, Mohanpur Campus, Nadia 741246, West Bengal, India}
\begin{abstract}
We describe a simple and accurate differential thermal analysis set up to measure the latent heat of solid state materials undergoing abrupt phase transitions in the temperature range from 77 K  to above room temperature. We report a numerical technique for the absolute calibration of the latent heat of the transition, without the need of a reference sample. The technique is applied to three different samples---vanadium sesquioxide undergoing the Mott transition, bismuth barium ruthenate undergoing a magnetoelastic transition, and an intermetallic Heusler compound. In each case, the inferred latent heat value agrees with the literature value to within its error margins. To further demonstrate the importance of absolute calibration, we show that the changes in the latent heat of the Mott transition in vanadium sesquioxide (\vo) stays constant to within 2\% even as the depth of supersaturation changes by about 10 K, in non-equilibrium dynamic hysteresis measurements. We also apply this technique for the measurement of the temperature-dependent specific heat.
\end{abstract}
\date{July 2021}
\preprint{\href{https://doi.org/10.1063/5.0056857}{{\it Review of Scientific Instruments} {\bf 92}, 075106 (2021).}}
\maketitle

\section{Introduction}
Differential thermal analysis (DTA) and differential scanning calorimetry (DSC) are widely used for the determination of the reaction heat, the specific heat, the latent heat, and the enthalpy in the temperature range from millikelvin to kilokelvin.\cite{Book_Springer, Book_Cambridge, Book_Royalsociety} While a variety of commercial DTA instruments are now available, a lack of confidence in the data has been expressed in many studies as the dependence of DTA peak's area on temperature scanning rate, heat flow rate, and transition temperature is not completely understood.\cite{Gmelin_PAC95, Hohne_acta89, Callanan_RSI87} Such extrinsic instrument-related effects may mask the small effects intrinsic to the samples, especially when the measurements are carried out under non-equilibrium conditions.

For example, while the latent heat is generally considered a material constant, thermally-induced hysteretic martensitic transformations are also thought to experience an increase in latent heat with the increasing scanning rate due to the irreversible motion of phase interface.\cite{Liu_SSC96, Wollants_PMS93, Kuang_SM00} Similarly, a small dependence of the latent heat of freezing of water on the supersaturation temperature has also been reported.\cite{Manosa_DSC} It is usually very difficult to unambiguously draw such inferences because the area of the latent heat peak in the DTA measurement itself varies strongly (approximately with a linear dependence) with the temperature scanning rate.\cite{Book_Springer, Book_Cambridge, Book_Royalsociety} Secondly, the DTA and DSC instruments are calibrated using well-established reference materials. The uncertainties in calibrant data, which are between 2-10\%, \cite{Gmelin_PAC95} always produce some errors.\cite{Ref_material} Since the calibration values are based on measurements supposedly under (quasi-)thermal equilibrium,\cite{Gmelin_PAC95, Ref_material} the difficulty is the interpretation of the measurements under non-equilibrium conditions is only compounded.

Despite the long history and its relative simplicity, there have been constant innovations in the methodology and instrumentation of DTA. Variants of different adiabatic and nonadiabatic calorimetry techniques---the heat pulse method,\cite{Wilhelm_rsi04, Gillespie_rsi20} the continuous heating method,\cite{Schilling_rsi07} the hybrid method,\cite{Klaasse_rsi08} the relaxation calorimetry,\cite{Cooke_rsi11} the heat flux calorimetry,\cite{Basso_rsi10, Razouk_rsi13, Gillespie_rsi20, Lacouture_rsi20} and the ac-calorimetry\cite{Miyoshi_rsi08, Morrison_rsi12}---have been discussed over the past 15 years.

Here we report a relatively simple and highly accurate lock-in amplifier-based DTA apparatus. The calibration is based on a first-principles technique where the absolute value of the latent heat can be estimated without the need for a reference sample. The robustness of the technique is further demonstrated by the fact that the calculated latent heat is found to vary less than 2\%, even as the temperature scan rate is varied by an order of magnitude and the DTA peak area changes by more than a factor of 10. Some of the measurements reported in references \onlinecite{Bar_prl18, Kundu_prl20} were done on this apparatus.

\section{DTA calorimeter}
\begin{figure}[!t]
\center
\includegraphics[scale=0.27]{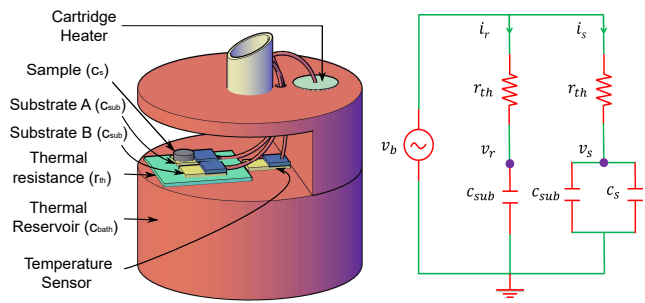}
\caption{(left) Schematic of the calorimeter. The sample is mounted on the substrate {\tt A} and the other nearly identical {\tt Pt-100} (substrate {\tt B}) serves as the reference. An external temperature controller controls the temperature of the thermal reservoir through a temperature sensor and a cartridge heater which is placed inside the thermal bath. (right) Equivalent electrical circuit of calorimeter based on analogy between the Biot-Fourier and the Ohm's law.}
\label{calorimeter}
\center
\end{figure}

Figure \ref{calorimeter} (left) shows the schematic of the calorimeter, which can be operated down to 77 K. Two nearly identical and calibrated {\tt Pt-100} resistors, in a ceramic package, serve as the substrates, {\tt A} and {\tt B} respectively, and are also used to measure the temperature.\cite{Nagapriya_prl06, Schilling_prb95, Schilling_rsi07} These two substrates are placed over a temperature-controlled copper block (heat capacity $c_{bath}\approx$ 31.74 J/K) that serves as a thermal bath. A glass coverslip of thermal resistance $r_{th}\approx$ 290 K/W  performs as a poor thermal link between the substrates and the thermal bath. The sample is mounted on the substrate {\tt A}. The substrate {\tt B} acts as the reference. When the latent heat is absorbed from (released to) the substrate {\tt A} during the transition, its temperature $T_A$ decreases (increases) with respect to the temperature $T_B$ of the reference substrate {\tt B}. Transition temperature and latent heat can be estimated from the temperature difference of two substrates ($T_A - T_B$). $T_B$ is also, of course, the temperature of the thermal environment experienced by the sample. The copper block can be thought of as a thermal reservoir because of its much larger heat capacity in comparison to the sample and the substrate heat capacities, $c_{s}$ and $c_{sub}$ respectively. It is critical to note that thermal resistance (coverslip) plays an important role by slowing the temperature equilibration of the substrate {\tt A} with the thermal bath. The temperature of the set up is controlled by a $Lakeshore$ temperature controller with the aid of a cartridge heater and another {\tt Pt-100} thermometer is mounted on the copper block (thermal bath). Good thermal contacts between the sample and the substrate, the substrates and the cover slip, etc. are ensured by a thin layer of suitable vacuum ($Apiezon\, N$) grease.

This calorimeter forms a small unit that can be placed inside most conventional cryostats. We have used a four-chamber liquid nitrogen cryostat (Oxford Instruments' $Optistat\,DN$) with a cold-finger design, that was rewired for the purpose. Static exchange gas was sometimes used to increase the cooling rate.

A nearly perfect linear temperature ramp between $77$-$300$ K at scan rates varying between $0.1$ K/min to about $50$ K/min could be accomplished \cite{Bar_prl18} by a suitable choice of the `proportional, integral, and derivative' (PID) values in the temperature controller.

\begin{figure}[b!]
\center
\includegraphics[scale=0.38]{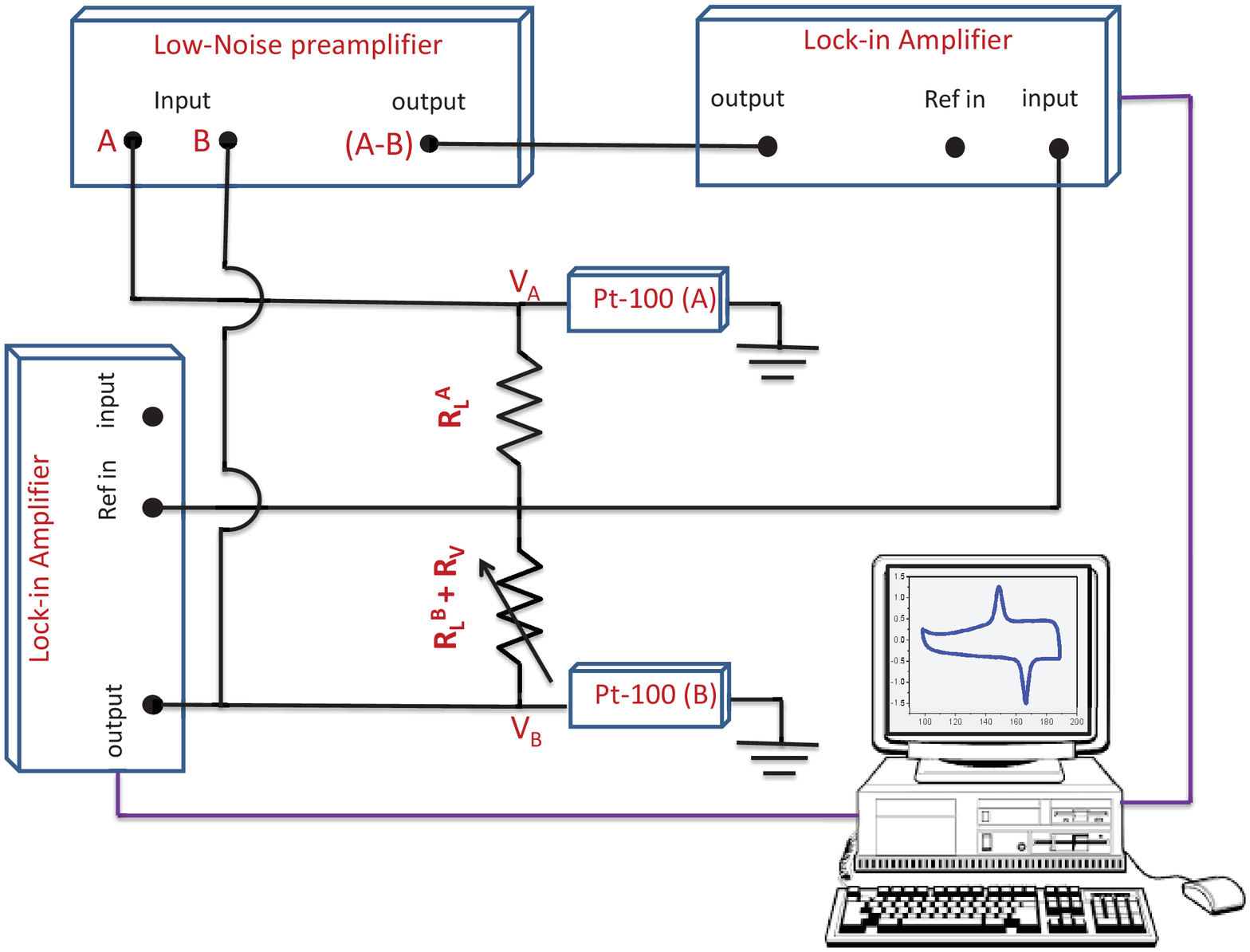}
\caption{Schematic of the electronics in the latent heat measurement setup. The resistance difference of two {\tt Pt-100} substrates {\tt A} and {\tt B} are measured with respect to the resistance of reference substrate {\tt B}.  Subsequently resistances are converted to the temperature from {\tt Pt-100} calibration data.}
\label{Measurement_setup}
\end{figure}

The electrical circuit with two lock-in amplifiers (LIA) ({\em Stanford Research Systems} model {\em SR 830}) is shown in Fig.\ref{Measurement_setup}, where one LIA measures the absolute temperature of the reference substrate {\tt B} and, the second LIA measures the temperature difference between the two substrates.  During the abrupt phase transition, as the latent heat is exchanged with the substrate {\tt A}, there is change in relative temperatures ($\triangle T$) between the substrates {\tt A} and {\tt B}. This DTA peak is shown in Fig. \ref{DTA_3Samples} as a function of reference temperature.

\section{Theory and calibration}
The assumption of linear response leads to the straightforward analogy between the heat transport and electrical circuit equations,\cite{Basso_rsi10, Lacouture_rsi20} where the amount of heat, the heat flux, the temperature difference,  the specific heat, the thermal resistance correspond respectively to an electrical charge $q$, electric current $i$, voltage difference $v$,  capacitance $c$, and electrical resistance $r$. We will use lowercase letters ($q$, $i$, $v$, $c$, $r$) to designate the electrical equivalents of the thermal circuit parameters and the actual electrical circuit parameters will be designated with the uppercase symbols ($R$, $I$, $V$, $C$).
The equivalent electrical circuit for the calorimeter in Fig.\ref{calorimeter} (left) is shown in Fig. \ref{calorimeter} (right). The equivalent circuits for the substrates {\tt A} and {\tt B} effectively decouple into parallel configurations under common bias with
\begin{equation} \label{eq:1}
v_s (c_{sub}+c_{s}) = \int i_s dt \ \ \& \ \ v_r c_{sub} = \int i_r dt,
\end{equation}
where
\begin{equation} \label{eq:2}
i_s = \frac{v_b-v_s}{r_{th}}\ \ \&\ \ i_r = \frac{v_b-v_r}{r_{th}}.
\end{equation}
Equation (\ref{eq:1}) can be solved numerically.

\begin{figure}[!t]
\center
\includegraphics[scale=0.34]{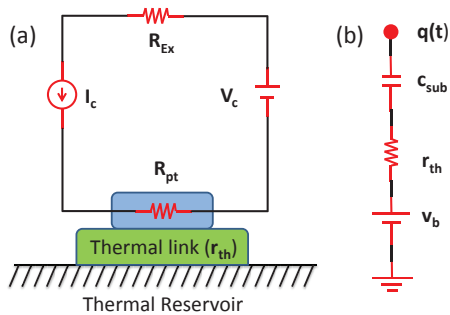}
\caption{Calibration:  Schematic of the (a) actual circuit, (b) electrical equivalent of the thermal circuit made for the determination of $c_{sub}(T)$ and $r_{th}(T)$.}
\label{Calibration_setup}
\center
\end{figure}

\begin{figure}[h!]
\center
\includegraphics[scale=0.34]{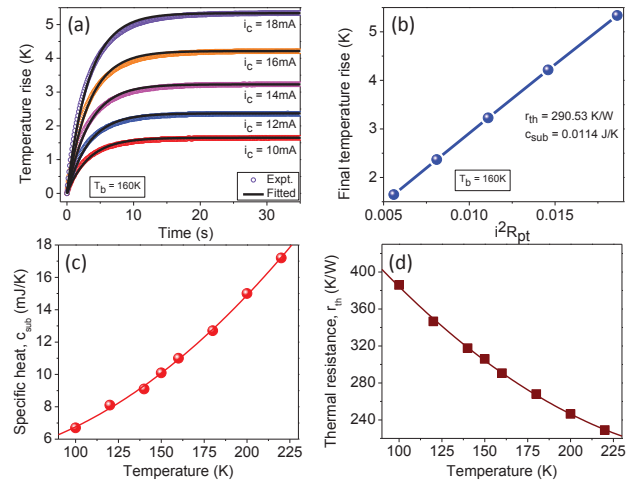}
\caption{(a) ($\circ$) Temperatures rise of the substrate due to continuous heating of {\tt Pt-100} by applying external currents (I$_c$= 10, 12, 14, 16, and 18 mA) at bath temperature T$_b$ = 160K. The Eq. \ref{ckt_response} are fitted with identical time constant $\tau = r_{th}c_{sub}\approx 3.2$ s (---). (b) Steady state temperatures ($\frac{I_c^2 R_{pt} \tau}{c_{sub}}$) $vs$ heating powers ($I_c^2 R_{pt}$) obey a straight line relationship, give rise to c$_{sub}$ = 0.011 J/K and r$_{th}$=290.53 K/W. Temperature dependent specific heat (c) of the substrate and thermal resistance (d) of the heat link are interpolated (line) form the repetitive calibration experiment at different steady-state temperatures (dots).}
\label{Calibration_data}
\center
\end{figure}

The material parameters $c_{sub}$ and $r_{th}$ are functions of temperature and their values (at any given temperature) are determined using the electric circuit shown in Fig. \ref{Calibration_setup}. Here a constant voltage $V_c$ is switched on as a step at time $t=0$ from an external power supply and the {\tt Pt-100} substrate is used to both dissipate the $I_c(t)^2 R_{pt}(t)$ heat per unit time, where $I_c(t)$ is the current through it and $R_{pt}(t)$ its resistance, as well as to monitor its temperature which increases to a steady state value that is $1-5$ K higher than the starting temperature $T_b$  [Fig \ref{Calibration_data} (a)].\cite{Cooke_rsi11} For a small rise in temperature, the change in current with time $I_c(t)$, due the change in {\tt Pt-100}'s resistance, can be ignored in the first-order approximation. Then the
temperature rise
\begin{equation} \label{ckt_response}
\Delta T(t) = {I_c^2 R_{pt} \tau \over c_{sub}} [1-e^{-t/\tau}]
\end{equation}
can be easily calculated from the electrical equivalent of the thermal circuit [Fig. \ref{Calibration_setup} (b)]. Here $\tau = c_{sub} r_{th}$ is the time constant. From Eq. \ref{ckt_response} it is evident that for sufficiently large values of $t$, the voltage reaches its steady-state value ${I_c^2 R_{pt} \tau}/{c_{sub}}$.

Figure \ref{Calibration_data} (a) shows the rise in {\tt Pt-100} substrate temperature with I$_c$= $10$, $12$, $14$, $16$, and $18$ mA and T$_b$ = $160$ K. The rise time $\tau$ is found to be $3.2$ s from electrical analogy of the thermal circuit (Fig. \ref{Calibration_setup}). The saturation temperature exhibits a linear dependence on thermal power (Fig. \ref{Calibration_data} (b)) with slope $\tau/c_{sub}$ giving $c_{sub}$ = 0.011 J/K and r$_{th}$ = 290.53 K/W.

The temperature dependence of $c_{sub}$ and $r_{th}$  is obtained by repeating the same experiment at a few different bath temperatures and a functional form is empirically determined by interpolating these few point with a polynomials of degree 2 [Fig. \ref{Calibration_data} (c), (d)].

\section{Results and analysis}
\subsection{DTA measurements and extraction of latent heat}
\begin{figure}[!b]
\center
\includegraphics[scale=0.34]{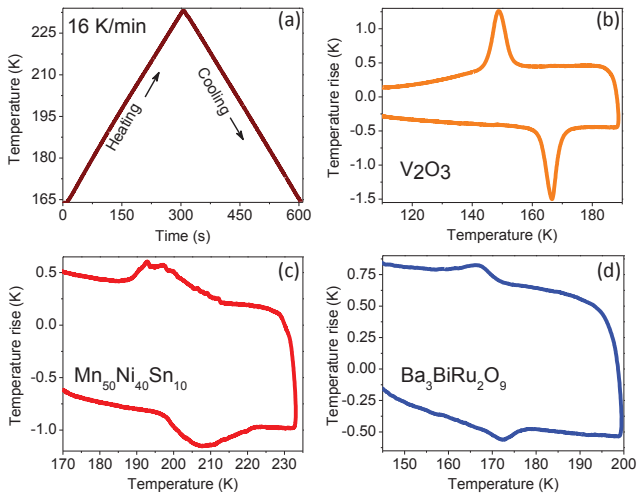}
\caption{(a) Illustrative temperature profile of a ramp signal which is the driving force of our experiments. DTA signal as a function of temperature for metal insulator transition of V$_2$O$_3$ (b), magneto-structural phase transition of  Mn-Ni-Sn based Heusler alloy (c), and  magnetoelastic transition of Ba$_3$BiRu$_2$O$_9$ (d). The temperature ramp rate of experiments was $16$ K/min.}
\label{DTA_3Samples}
\center
\end{figure}

Figure \ref{DTA_3Samples} shows DTA measurements for abrupt phase transitions in three different materials. The DTA peaks\cite{Lh_V2O3} of V$_2$O$_3$ [Fig. \ref{DTA_3Samples} (b)]  are quite sharp indicating an abrupt phase transition (APT) with a relatively large latent heat and the single phase nature of the material.\cite{Bar_prl18} The latent heat peaks associated with the magnetoelastic transition in Ba$_3$BiRu$_2$O$_9$ and martensitic transformation in MnNiSn alloy, on the other hand, are relatively small.\cite{Lh_Ba3BiRu2O9, Lh_MnNiSn} The noise in the DTA peaks in Fig. \ref{DTA_3Samples} (c) indicates avalanches \cite{Toth_prb14} during the martensitic transition and is not a measurement limitation. The overall shape of the DTA curve away from the transition is determined by temperature dependence the heat capacity of the sample.\cite{Schilling_prb95}

\begin{figure}[b!]
\center
\includegraphics[scale=0.34]{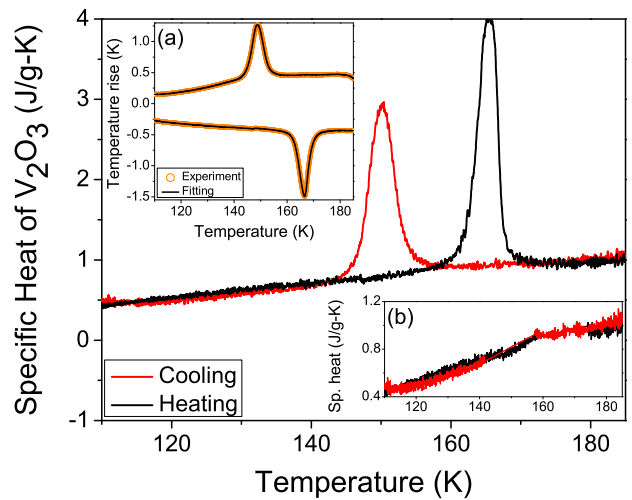}
\caption{(inset a) Temperature-dependent specific heat of V$_2$O$_3$ is inferred by varying (only) the sample heat capacity $c_s(T)$ in the equivalent circuit [Fig \ref{calorimeter} (right)] such that the simulated signal fits the experimental DTA graph. (inset b) The contribution to the DTA peaks at the transition region is primarily from the latent heat of transition and the values of the specific heat in this region are inferred by linearly extrapolating the sample specific heat outside this region. The main figure shows the inferred specific heat of the sample in the complete temperature range. The difference in the regions around the transition is attributed to the latent heat which is added in the next round of the circuit simulation. }
\label{Spheat}
\center
\end{figure}

Since the values of the thermal circuit parameters $c_{sub}$ and $r_{th}$ have already been determined, the curves observed in Fig. 5 are easily simulated by numerically integrating the equivalent circuit equations with a linearly ramped temperature. Note that the signals in Fig 5 (b, c, d) are the measurements of the temperature difference between the sample-containing substrate {\tt A} and the blank substrate {\tt B}. The temperature dependence of this difference can arise from two factors---(a) the temperature dependent changes in the heat capacity of the sample, or (b) the release or absorption of the latent heat by the sample. Unfortunately, these two factors are not clearly separable and one may theoretically simulate the observed curves by either varying the sample heat capacity $c_s$, or (ii) introducing or eliminating a certain amount of charge on $c_s$ to simulate the effect of the release or the absorption of the latent heat. While there is no unambiguous way of dealing with this problem, one way to systematically proceed is the following: We first assume that there is no latent heat contribution and simulate the particular temperature rise $\Delta T(T)$ curve by varying only the heat capacity of the sample, $c_s(T)$, in the equivalent circuit [Fig. \ref{calorimeter} (right)]. One such sample calculation is shown in Fig. \ref{Spheat} (inset a). Secondly, we note that while the temperature variation of specific heat of V$_2$O$_3$ is in good agreement with the literature values \cite{Heat_capacity} {\it away} from the transition, the peaks in Fig. \ref{Spheat} are not physical as indeed the specific heat is not defined during at the APT. The DTA curve around the APT is governed by the latent heat of the transition. To account for both these facts, we have demarcated regions in the DTA curve which show a steep changes in the slope (due to the latent heat release at the APT) from the smoothly varying background. This region of steep increase was now eliminated from the analysis and the specific heat values $c_s(T)$ in the region of the phase transition were linearly interpolated from the values of the specific heat just before and just after the transition [Fig. \ref{Spheat} (inset b)].

This linear interpolation of $c_s(T)$ [Fig. \ref{Spheat} (down-inset)] in the region of the abrupt phase transition is obviously unphysical and is meant to only give a handle on the background of the latent heat peak. Since the substrate's specific heat [Fig. \ref{Calibration_data}(c)] and thermal resistance [Fig. \ref{Calibration_data}(d)] in the equivalent circuit [Fig. \ref{calorimeter} (right)] were already estimated, we can finally fit the DTA peak by externally adding (or extracting) a certain amount ($q$) of heat in the circuit simulation at each time step. The integrated value of this assigned heat is the latent heat of transition. The latent heat of \vo is listed in table \ref{table}. The same idea was implemented to calculate the latent heat of Ba$_3$BiRu$_2$O$_9$ and  Mn-Ni-Sn alloy. The present experimental results of three samples agree within 5\% with the literature values which themselves have similar uncertainties.
\begin{table}[b]
\caption{Estimates of the latent heat (during heating) for three different samples undergoing hysteretic abrupt transitions.}
\label{table}
\begin{tabular}{p{2.1cm}p{1.5cm}p{1.8cm}p{2.7cm}}
\hline
\hline
Samples & Mass & Latent Heat & Literature value \\
\hline
V$_2$O$_3$ & 6.59 mg & 1650 J/mol & 1490-2100 J/mol \cite{Lh_V2O3}\\
Mn$_{50}$Ni$_{40}$Sn$_{10}$ & 7.76 mg & 8.47 J/g & 6.63-12.42 J/g \cite{Lh_MnNiSn}\\
Ba$_3$BiRu$_2$O$_9$ & 14.8 mg & 715 J/mol &  710-720 J/mol \cite{Lh_Ba3BiRu2O9}\\
\hline
\hline
\end{tabular}
\end{table}

We must mention that there is some arbitrariness in the choice of the transition region. Since the change prominence of the latent heat peak over the smooth specific heat background is strongly dependent on the temperature scanning rate $R$, our technique does not work for rates $R \leq $ 1 K/min.  In the case of a small ramp rate, a small error in locating those points produces a large error in the latent heat value as background subtraction of the latent heat peak has been made depending upon those points. Due to the large size of the DTA peak in the high ramp rate, the background correction does not affect much on the latent heat values.
\begin{figure}[h!]
\center
\includegraphics[scale=0.4]{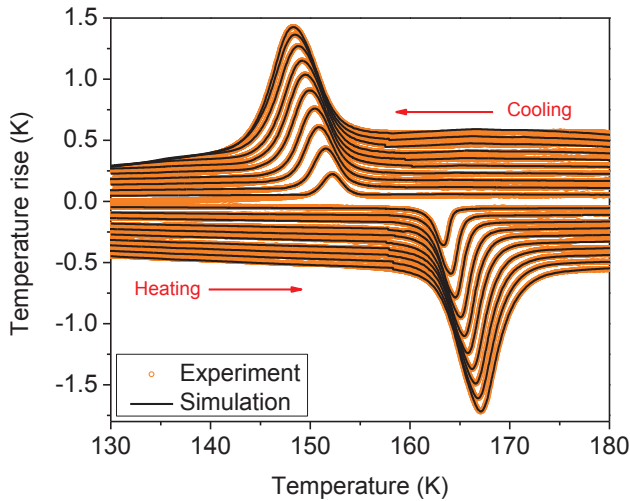}
\caption{($\circ$) DTA signal as a function of temperature for different linear ramp rates. (---) The simulated DTA signals. In each case, the values of the free system parameters, $c_s(T)$, $c_{sub}$, and $r_{th}$ are the same.}
\label{DTA_fits}
\center
\end{figure}

\begin{figure}[h!]
\center
\includegraphics[scale=0.4]{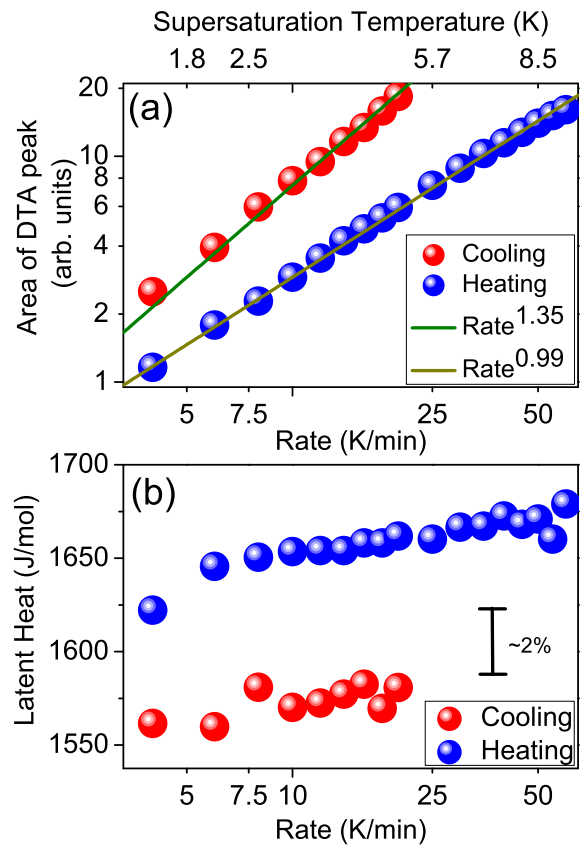}
\caption{(a) Change in DTA peak's area (arbitrary unit) with temperature quench rates. (b) Latent heat of \vo with the depths of supercooling and superheating shift (top X-axis) due to different driven temperature scanning rate (bottom X-axis). The supersaturated temperature ($\Delta T$) follows a scaling law with scanning rate (R): $\Delta T \sim  R^{2/3}$.}
\label{Latent_heat}
\center
\end{figure}
\subsection{Latent heat of supersaturated transition}
Based on the analysis of the data in Fig. 7, we had previously observed that the depth of supercooling and superheating increases with the temperature scan rate $R$, and follows a dynamical scaling relation $\Delta T(R)\propto R^\Upsilon$.\cite{Bar_prl18} Here $\Delta T(R)=|T_0^{j}-T^{j}_{R}|$, $j=H, C$,  is the shift in the phase transition temperature to an $R$-dependent value $T^H_{R}>T^H_0$ during heating and $T^C_{R}<T^C_0$ during cooling. $T^H_0$ and $T^C_0$ are the transition temperature under a quasistatic change in the sample temperature under heating and cooling conditions respectively.

It is natural to ask if the value of the latent heat would also be scan rate-dependent. One may naively expect that the sample's heat loss or gain would be proportional to the area of the curves measuring the differential temperature rise [Fig. 7].\cite{Liu_SSC96} But the primary rate dependence of the temperature rise, for intermediate temperature scan rates $R$, actually comes from the competition between the rate of the release of the latent heat and the thermal time constant $\tau=r_{th}[c_{sub}+c_{s}]$ of the apparatus. This is, of course, a well-known fact.\cite{Book_Springer, Book_Cambridge, Book_Royalsociety, Kuang_SM00, Schilling_prb95} The actual change in the DTA peak area with $R$ is shown in Fig. \ref{Latent_heat} (a), for over an order of magnitude variation of the temperature scan rate.

In Fig. \ref{DTA_fits}, the experimental DTA curves for different temperature scan rates are fitted to estimate the latent heat of V$_2$O$_3$. The values of the free parameters $c_s(T)$, $c_{sub}(T)$, and $r_{th}(T)$ are kept fixed at their previously determined values. It is perhaps remarkable that despite more than a 1000\% change in the area of the DTA curve [Fig. 8 (a)], the inferred latent is found to be constant with about 2\% [Fig. 8 (b)]. The difference in latent heat in heating and cooling is usually attributed to the extra dissipation caused by internal-friction.\cite{Zhang_prb95, Ortin_acta88}

\section{Conclusion}
We have implemented an absolute calibration technique in a homemade DTA set up. This technique therefore does away with the need of a standardized calibrant sample. The reliability of the apparatus is demonstrated by latent heat measurements in three different materials---vanadium sesquioxide undergoing the Mott transition, bismuth barium ruthenate undergoing a magnetoelastic transition, and an intermetallic Heusler compound. In each case, the value of latent heat determined by this method agrees very well with the reported data. It was shown that, despite a very large temperature sweep rate-dependence of the DTA signal, where the area of the DTA peak changed by about 2000\%, the value of the inferred latent heat is constant to within about 2\%. The technique is shown to be capable of also reliably computing the specific heat of the samples at the same time.

The technique described here can be easily extended to much lower (up to $\sim 4$ K) or higher temperatures by appropriate choice of the sensors and the cryostat. The compact apparatus can also easily be accommodated within one of the standard variable temperature inserts of, for example, a `physical property measuring system' to enable magnetic field-dependent measurements.
\section{Acknowledgments}
We would like to thank K. S. Sujith for help to design the experimental setup, Dr. Sunil Nair, Dr. Arup Ghosh, and Dr. Satyabrata Raj for providing samples, Mr. Pintu Das for technical support.

\section{Data Availability}
The raw data and the computer programs are available upon request.

\end{document}